\documentclass[]{ceurart}
\usepackage{caption}
\usepackage{subcaption}
\usepackage{multirow}
\usepackage{tabularx}
\usepackage[font=small,skip=2pt]{caption}
\usepackage{listings}
\begin{document}

\copyrightyear{2023}
\copyrightclause{Copyright for this paper by its authors.
  Use permitted under Creative Commons License Attribution 4.0
  International (CC BY 4.0).}

\conference{IntRS'23: Joint Workshop on Interfaces and Human Decision Making for Recommender Systems, September 18, 2023, Singapore (hybrid event)}

\title{Concentrating on the Impact: Consequence-based Explanations in Recommender Systems}


\author[1]{Sebastian Lubos}[%
email=slubos@ist.tugraz.at
]
\author[1]{Thi Ngoc Trang Tran}[%
email=ttrang@ist.tugraz.at
]
\author[1]{Seda Polat Erdeniz}[%
email=sedapolat@gmail.com
]
\author[1]{Merfat El Mansi}[%
email=merfat.el-mansi@student.tugraz.at
]
\author[1]{Alexander Felfernig}[%
email=afelfern@ist.tugraz.at
]
\author[2]{Manfred Wundara}[%
email=manfred.wundara@villach.at
]
\author[3]{Gerhard Leitner}[%
email=gerhard.leitner@aau.at
]
\address[1]{Institute of Software Technology, Graz University of Technology,
  Inffeldgasse 16b, 8010 Graz, Austria}
\address[2]{Magistrat Villach,
  Rathausplatz 1, 9500 Villach, Austria}
\address[3]{Interactive Systems Group, University of Klagenfurt,
  Universitätsstrasse 65-67, 9020 Klagenfurt am Wörthersee, Austria}



\begin{abstract}
Recommender systems assist users in decision-making, where the presentation of recommended items and their explanations are critical factors for enhancing the overall user experience. Although various methods for generating explanations have been proposed, there is still room for improvement, particularly for users who lack expertise in a specific item domain. In this study, we introduce the novel concept of \textit{consequence-based explanations}, a type of explanation that emphasizes the individual impact of consuming a recommended item on the user, which makes the effect of following recommendations clearer. We conducted an online user study to examine our assumption about the appreciation of consequence-based explanations and their impacts on different explanation aims in recommender systems. Our findings highlight the importance of consequence-based explanations, which were well-received by users and effectively improved user satisfaction in recommender systems. These results provide valuable insights for designing engaging explanations that can enhance the overall user experience in decision-making.
\end{abstract}

\begin{keywords}
  recommender systems \sep
  consequence-based explanations \sep
  decision-making \sep
  human-centered computing
\end{keywords}

\maketitle

\section{Introduction} \label{sec:introduction}
\textit{Recommender systems (RS)} assist users in decision-making by predicting which items in a catalog will be of interest to them \cite{ricci2022recommender}. They help users to make efficient and satisfying decisions with many available options in various domains, including, movie selection \cite{gomez-uribe2015netflix,sharma2020movie}, meal choices \cite{ge2015health,pecune2020recommender}, or apartment considerations \cite{burke2000knowledge,gharahighehi2021recommender}. The presentation of recommendations is crucial to the overall user experience, and explaining them is an important aspect \cite{bilgic2005explaining}. Although current explanation approaches have proven useful to extend RS, the need for improvements remains, specifically in terms of generating natural language explanations and explaining impacts \cite{zhang2020explainable}.

To address these needs, we introduce \textit{consequence-based explanations}, which focus on the impact of consuming recommended items as the main argument. These explanations are particularly beneficial in domains where users have limited knowledge and struggle to understand the effects of following recommendations. By emphasizing the actual changes and outcomes resulting from a recommendation, users can gain a clearer understanding of how their choices directly influence their circumstances. To validate this concept, we conducted an online user study to assess users' appreciation of consequence-based explanations and compared different ways of formulating them in two domains with different levels of involvement \cite{felfernig2017analysis}. Our study provides initial guidance for generating consequence-based explanations.

\section{Related Work} \label{sec:related_work}
Explaining recommendations and providing users with an understanding of why a particular item is recommended  emerged as an important topic in recent years \cite{zhang2020explainable}.  Providing such explanations is critical for shaping the user experience of an RS \cite{bilgic2005explaining}. Consequently, seven explanation aims have been suggested \cite{tintarev2007survey}: (i) \textit{Effectiveness} to help users make good decisions, (ii) \textit{Efficiency} to help users make decisions faster, (iii) \textit{Persuasiveness} to convince users, (iv) \textit{Satisfaction} to improve the user experience, (v) \textit{Scrutability} to allow users to correct the system, (vi) \textit{Transparency} to explain how the system works, and (vii) \textit{Trust} to increase users' confidence in the system. Studies investigating the effect of different methods for generating and visualizing explanations show that there is no optimal approach that fits all dimensions, and explanations should be tailored to the specific goal of the RS \cite{gedikli2014should}. 

Explanations in RS influence users by persuading them to consume recommended items \cite{gkika2014persuasive}, increasing the perceived usefulness of the RS \cite{zanker2010knowledgable}, and contributing to overall user satisfaction \cite{tran2021users}. Domain-specific content data enhances explanation effectiveness, while better transparency leads to higher user satisfaction \cite{gedikli2014should}. Users prefer familiar explanation types, as those reduce cognitive effort. However, pure optimization for efficiency is not always useful, as users are willing to spend time analyzing explanations to make good decisions \cite{gedikli2014should}.

Explanation types can be categorized depending on how they are generated \cite{zhang2020explainable}. \textit{Model-intrinsic} approaches use interpretable recommendation models that directly provide explanations, while \textit{model-agnostic} approaches generate explanations from recommended items. One common technique at this level is the \textit{content-based} style explanation, which utilizes item features and past consumption data for generation \cite{tintarev2022beyond}. 

When users have difficulties evaluating recommended options due to limited domain knowledge, mentioning the impact of potential choices explicitly in a \textit{consequence-based} explanation can help. While this explanation type is new to RS, it has proven to be useful in other applications. Reinforcement learning agents explained their behavior in terms of expected consequences of state transitions, to increase human understanding and to enable evaluation of the plausibility of the agents' decisions \cite{vanderwaa2018contrastive}. Including potential consequences in tornado warnings increased the likelihood of persons taking protective actions \cite{ripberger_2015_influence}, and visualizing personal consequences of decisions supported financial planning \cite{fano2003personal}.

\section{Consequence-based Explanations} \label{sec:consequence_based_explanations}
In this paper, we introduce \textit{Consequence-based} explanations, which highlight the positive and negative impacts of consuming a recommended item. This type of explanation emphasizes the individual effects of choice rather than the underlying factors that led to it. They support users in decision-making, by highlighting how the decision toward an option influences the circumstances in a desired or undesired way. For example, a movie recommender might generate a consequence-based explanation like: \textit{"Toy Story has been recommended to you because it will entertain your whole family, and teach your children about the value of friendship"}. Those explanations can be created using a model-agnostic approach, using the feature descriptors of the recommended items, e.g., the genre and keyword tags. To simplify the decision-making process, it is advisable to include only important consequences for users, considering the multitude of potential outcomes. This aligns with how people typically communicate explanations, focusing on relevant causes rather than overwhelming with excessive information \cite{miller2019explanation}.

This paper explores explanations in two domains representing different \textit{levels of involvement} \cite{felfernig2017analysis}. While we examine \textit{apartments} as high-involvement items, requiring significant time and effort for decision-making due to their long-term and financial impact, we consider \textit{recipes} as a low-involvement domain with shorter-term and less serious impact\footnote{Unhealthy nutrition can have significant consequences, but the impact of a single decision to cook a recipe is low-involvement.}.

Consequences are derived using a rule-based approach customized to each domain, leveraging item features and user preferences (see Section \ref{subsubsection:dataset}). A sentence is prepared to explain the consequence of each feature value associated with a recommended item, taking into account the specified user preferences. These individual sentences are then combined to form the complete explanation for the recommendation. For instance, the recipe recommender considers user data such as \textit{activity level} and \textit{weight aim} to suggest a recipe with suitable nutrient quantities. By integrating these preferences with nutritional data of recipes, the appropriate sentence for the overall explanation is determined.

We distinguish two types of consequence-based explanations based on their formulation: (i) \textit{motivating consequence-based} explanations formulate the impact in a positive way, expressing which favorable consequence the suggested item has (ii) \textit{avoiding consequence-based} explanations highlight the negative impact that can be prevented by choosing the suggested item. An example for both types is presented in Table \ref{tab:explanation_examples}. To ensure transparency, the downsides of a recommendation are also communicated, if the suggested item does not fulfill all user preferences. For instance, in the apartment domain, an explanation might state that "\textit{your children will need to share bedrooms in this apartment}".

\begin{table}[t]
  \caption{Examples of \textit{motivating} and \textit{avoiding} consequence-based explanations, using predefined sentences. The motivating formulation emphasizes the positive impact of consuming the recommended recipe on the users' activity level and weight aim, while the avoiding formulation highlights the negative impacts on those properties that are prevented with the recommended recipe.}
  \label{tab:explanation_examples}
  \begin{tabular}{p{0.47\textwidth}p{0.48\textwidth}}
    \toprule
    \textbf{Motivating Consequence-based} &   \textbf{Avoiding Consequence-based} \\ \hline
    The number of carbs, sugar, and protein in the cooked meal will \textbf{give you enough energy} for your activity level, and the number of calories and fat in the dish will \textbf{support you in losing weight}.  &   The number of carbs, sugar, and protein in the cooked meal will \textbf{not fall below the needed energy} for your activity level, and the number of calories and fat in the dish will \textbf{not interfere with your aim of losing weight}. \\
  \bottomrule
\end{tabular}
\end{table}
\raggedbottom

\section{Research Questions and User Study} \label{sec:research_questions_and_user_study_design}
This study evaluates two formulations of consequence-based explanations and their impact on explanation aims in RS \cite{tintarev2007survey}. It compares them to standard content-based explanations as the baseline across two domains with varying levels of involvement. The results offer initial guidance for generating consequence-based explanations that emphasize decision impact. The study is designed and conducted to answer four research questions:

\textbf{RQ1}: How do consequence-based explanations contribute to the explanation aims in RS?

\textbf{RQ2}: How do different formulations of consequences influence their contribution towards the explanation aims?

\textbf{RQ3}: How does the level of involvement of items affect the contribution of consequence-based explanations?

\textbf{RQ4}: Is there a correlation between users' demographics and preferences and the effect on the explanation aims?

Answers to those questions provide insights into (i) user appreciation, (ii) preferences, and (iii) the relevance of consequence-based explanations across domains. This helps improve user support in RS with more valuable explanations.

\subsection{User Study}
\subsubsection{Dataset Preparation} \label{subsubsection:dataset}
To prepare the user study, we created small datasets consisting of 20 items per domain. These datasets were designed with feature descriptors that included the necessary properties for generating explanations. For the recipe domain, data from recipe websites\footnote{https://www.aheadofthyme.com, https://www.seriouseats.com, https://www.themediterraneandish.com, and https://theplantbasedschool.com} was collected and processed in a uniform format including \emph{title}, \emph{description with ingredients}, and a \emph{generated image}\footnote{The photo was created with DALL-E 2, an AI image generator (\textit{https://openai.com/product/dall-e-2})} using the recipe title as input. Features include the \textit{cuisine type}, \textit{difficulty}, \textit{diet}, \textit{cooking time}, \textit{nutritional data}, and \textit{allergenic}.
For the apartment dataset, \textit{size}, \textit{rent}, \textit{number of bedrooms}, \textit{distance to the city center}, \textit{availability of private parking and garden}, and \textit{distance to leisure facilities} are used as features. Those samples have been created such that the content feature values are equally distributed. Those have been extended with a \emph{generated photo}, \emph{title}, and \emph{description} in the style of an apartment advertisement\footnote{Descriptions were generated with ChatGPT using features as input. Their correctness has been inspected manually.}.

\subsubsection{Generation of Recommendations} \label{subsubsection:generation_recommendations}
Since this paper focuses on explanations, the implemented RS is simple but generates personalized recommendations based on user preferences and item features. Preferences in the recipe domain, include \textit{favorite cuisine}, \textit{followed diets}, \textit{preferred cooking time}, \textit{cooking skill}, \textit{ingredients to be avoided}, \textit{activity level}, and \textit{weight aim}. In the apartment domain, preferences include the \textit{number of children living in the apartment}, \textit{rent}, \textit{preferred distance to the city center}, \textit{availability of a private car}, and \textit{favorite leisure time activities}. 
Recommendations are generated in a two-step process of candidate generation and ranking \cite{davidson2010youtube}. Firstly, the dataset is filtered for candidates fulfilling strict user preferences, i.e., diet, ingredients, cooking time, and cooking skill for the recipe domain, and city center distance and rent for the apartment domain. The remaining candidates are then ranked by compatibility with the \textit{Multi-Attribute Utility Method (MAUT)} \cite{jansen2011maut}, assigning scores to item alternatives, depending on their overlap with feature preferences.  

\subsubsection{Study Design}
We conducted an online study adopting the design used in \cite{bilgic2005explaining}. Participants were randomly assigned to either of the domains, where they were presented with a scenario of deciding on a recipe to cook later that day or searching for an apartment to move in for the next years, based on their personal situation and interests. After providing demographic information (age, gender, education) and preferences (see Section \ref{subsubsection:generation_recommendations}), a suggested item was presented, without any content information, only described by the generated explanation. Participants were evenly distributed among the different explanation variants (motivating or avoiding consequence-based and content-based). 
They used a \textit{5-point Likert scale} to rate their likelihood of choosing the suggested item, along with their perception of the explanation's satisfaction and understandability. Furthermore, the importance of individual consequences was rated to evaluate suggestion quality. In the final step, participants received another suggestion with a complete item description and rated their likelihood of considering it. Notably, participants were unaware that the same item was suggested twice with different descriptions, enabling the assessment of explanation effects by comparing rating differences.

\section{Evaluation} \label{sec:evaluation}
\subsection{Study Participants}

In total, $114$ persons participated in the study. $63$ were part of the recipe domain, while $51$ used the apartment recommendation. Around $25\%$ of participants identified as female. In general, the participants were rather young, with an average of $26$ years. Influenced by this, $51\%$ of the participants specified \textit{high school} as their highest educational degree, while $45\%$ had graduate degrees. These demographic factors need to be considered to interpret the results.

\subsection{Analysis Methods}
After gathering participant responses, we evaluated the impact of consequence-based explanations on the explanation aims \cite{tintarev2007survey} using metrics proposed in \cite{gedikli2014should}, comparing the results to a content-based explanation baseline.
The collected datasets share common features, such as non-normal distribution\footnote{following the \textit{Shapiro-Wilk} test, with $\alpha < 0.05$}, contain ordinal values, and violate independence due to the within-subjects study design. Therefore, non-parametric tests were used in \textit{SPSS} as hypothesis tests with $\alpha = 0.05$. In case the hypothesis test showed significance, the value of $\alpha$ was updated for pairwise follow-up tests as $\alpha_{new} = \frac{\alpha}{N}$, where $N$ is the number of pairwise follow-up tests of the respective attributes, to avoid error type 1.

\subsection{Explanation Aims}
The user study aims to investigate the impact of consequence-based explanations on explanation aims in RS \cite{tintarev2007survey}, comparing different formulations in domains with varying levels of involvement. It assesses efficiency, effectiveness, persuasiveness, satisfaction, and transparency, but does not examine the impact on trust and scrutability.

\subsubsection{Efficiency}
The study analyzed the \textit{item-based efficiency}, i.e., the time users spend to evaluate an item and decide on the likelihood of consuming it \cite{gedikli2014should}, of consequence-based explanations compared to content-based explanations in low and high-involvement domains, using recipes and apartments as representatives. The results (see Table \ref{tab:rating_statistics}) indicate that consequence-based explanations were more efficient, as users took less time to decide, especially with motivating consequences. The \textit{Kruskal-Wallis} test and follow-up \textit{Mann-Whitney-U} tests confirmed that those were significantly more efficient than avoiding consequences and the baseline in the recipe domain. However, there was no significant difference between avoiding consequences and content-based explanations or between explanation variants in the apartment domain. The better efficiency of consequence-based explanations suggests that including individual impact in the explanation helps users decide faster. The slight advantage of motivating consequences could be due to users' interest in the positive effects of a decision or lower cognitive effort if negations are avoided.

Comparing the domains, the item-based efficiency was slightly better in the apartment domain, without statistical significance. The duration differences between the motivating and avoiding consequence formulation were small within the domains. Avoiding consequences were more efficient in the apartment domain, likely because pointing out the serious avoided negative consequences helps users make faster decisions. In contrast, in the recipe domain, motivating consequences were more efficient, probably because the positive impact of consuming food is more relevant in this case. 

Overall, the analysis shows that consequence-based explanations improve decision-making efficiency. Considering, that they were longer ($338$-$749$ characters) compared to the content-based explanations ($186$-$524$ characters), the effect on effectiveness is further strengthened. The motivation of positive consequences seems to be more efficient in low-involvement domains, while the avoidance of negative impact tends to be more efficient in high-involvement domains. This aligns with the \textit{prospect theory} \cite{kahneman1979prospect}, indicating that people are risk-averse with gains but risk-seeking with losses. People tend to avoid loss in important decisions.

\begin{table}[t]
\caption{Mean study results by explanation aims \cite{tintarev2007survey}. The efficiency results represent the average time it took users to decide the likelihood of consuming an item based on the explanation. Effectiveness is calculated as the difference between the ratings given by a user for the explanation and the content description. A value close to zero indicates effectiveness, while a positive value suggests persuasiveness. For satisfaction and transparency, mean values of user-perceived ratings are shown.}
\begin{subtable}[]{\textwidth}
    \centering
    \begin{tabularx}{\textwidth}{>{\hsize=.44\hsize\linewidth=\hsize\centering\arraybackslash}X | >{\hsize=.13\hsize\linewidth=\hsize\centering\arraybackslash}X | >{\hsize=.13\hsize\linewidth=\hsize\centering\arraybackslash}X |>{\hsize=.13\hsize\linewidth=\hsize\centering\arraybackslash}X | >{\hsize=.13\hsize\linewidth=\hsize\centering\arraybackslash}X | >{\hsize=.13\hsize\linewidth=\hsize\centering\arraybackslash}X |>{\hsize=.13\hsize\linewidth=\hsize\centering\arraybackslash}X | >{\hsize=.13\hsize\linewidth=\hsize\centering\arraybackslash}X | >{\hsize=.13\hsize\linewidth=\hsize\centering\arraybackslash}X |>{\hsize=.13\hsize\linewidth=\hsize\centering\arraybackslash}X | >{\hsize=.13\hsize\linewidth=\hsize\centering\arraybackslash}X | >{\hsize=.13\hsize\linewidth=\hsize\centering\arraybackslash}X |>{\hsize=.13\hsize\linewidth=\hsize\centering\arraybackslash}X }
    \hline
    & \multicolumn{3}{c|}{Motivating} & \multicolumn{3}{|c|}{Avoiding}   & \multicolumn{3}{|c|}{Content-Based}    & \multicolumn{3}{|c}{Overall}   \\\hline
    Domains & Apt   & Rec & All & Apt   & Rec & All & Apt   & Rec & All & Apt   & Rec & All    \\\hline
    Efficiency [s]   &    132.9   &   130.5   &    131.6  &   125.7   &   138.8   &   133.0   &   173.1   &  147.5  &  159.3  &  143.9 & 138.8  & 141.1 \\\hline
    Effectiveness    &    0.4   &   0.1   &    0.2  &   0.2   &   -0.3   &   -0.1   &   0.2   &  -0.6  &  -0.3  &  0.2 & -0.3  & -0.1 \\\hline
    Satisfaction     &    3.8   &   3.4   &    3.6  &   3.8   &   3.7   &   3.7   &   3.5   &  3.3  &  3.4  &  3.7 & 3.5  & 3.6 \\\hline
    Transparency      &    3.9   &   3.6   &    3.8  &   3.8   &   3.7   &   3.7   &   3.8   &  4.0  &  4.0  &  3.8 & 3.8  & 3.8 \\\hline  
    \end{tabularx}
    \caption{\textbf{Explanation types and formulations}, grouped by domains. }
    \label{tab:rating_statistics}
\end{subtable}
\begin{subtable}[]{\textwidth}
    \centering
    \begin{tabularx}{\textwidth}{ >{\hsize=.235\hsize\linewidth=\hsize\centering\arraybackslash}X | >{\hsize=.125\hsize\linewidth=\hsize\centering\arraybackslash}X | >{\hsize=.125\hsize\linewidth=\hsize\centering\arraybackslash}X |>{\hsize=.125\hsize\linewidth=\hsize\centering\arraybackslash}X | >{\hsize=.125\hsize\linewidth=\hsize\centering\arraybackslash}X | >{\hsize=.125\hsize\linewidth=\hsize\centering\arraybackslash}X |>{\hsize=.125\hsize\linewidth=\hsize\centering\arraybackslash}X | >{\hsize=.125\hsize\linewidth=\hsize\centering\arraybackslash}X | >{\hsize=.125\hsize\linewidth=\hsize\centering\arraybackslash}X |>{\hsize=.125\hsize\linewidth=\hsize\centering\arraybackslash}X | >{\hsize=.125\hsize\linewidth=\hsize\centering\arraybackslash}X | >{\hsize=.125\hsize\linewidth=\hsize\centering\arraybackslash}X |>{\hsize=.125\hsize\linewidth=\hsize\centering\arraybackslash}X | >{\hsize=.125\hsize\linewidth=\hsize\centering\arraybackslash}X | >{\hsize=.125\hsize\linewidth=\hsize\centering\arraybackslash}X}
    \hline
    & \multicolumn{2}{c|}{Gender} & \multicolumn{2}{|c|}{Education}   & \multicolumn{3}{|c|}{Cooking Time}    & \multicolumn{4}{|c|}{Activity Level}& \multicolumn{3}{|c}{Weight Aim}    \\\hline
    Groups & F & M & Non-uni. & Uni.    &   No Pref.   & <30 min   &   <1h &   Sed.   &   Light  &   Mod.   &   Very &   Lose &   Mnt. &   Gain \\\hline
    Eff. [s]   & 147.5   &   133.2   &    126.9  &   146.4   &   165.6   &   132.1   &   124.3   &  143.9  &  126.0  &  128.0 & 183.4  & 131.2  & 151.2  &  121.0 \\\hline
    Effect.  &  -1.2    &   0   &   -0.2    &   -0.4   &  -0.9  & 0.1 & -0.3 & -0.4 & -0.4  &  0.2 &  -1 & -0.6  & -0.2  & 0.4  \\\hline
    Sat.   & 3.6 &   3.4 &   3.4 &   3.6   &  3.7  &  3.5  &  3.4 & 3.7 & 3.5  & 3.2  &  3.7 &  3.7 & 3.3  & 3.7  \\\hline
    Transp.   & 4.0   &   3.7  &  3.5  &  4.1 &  3.6  & 3.8  &  3.8  & 3.6 &  4.1 &  3.4  &  4.3 &  3.9 & 3.7  &  4.0\\\hline    
    \end{tabularx}
    \caption{\textbf{Low-involvement} domain (recipes), grouped by demographics and preferences (excluding \textit{preferred diet} due to even distribution).}
    \label{tab:recipe_demographics}
\end{subtable}
\begin{subtable}[]{\textwidth}
    \centering
    \begin{tabularx}{\textwidth}{ >{\hsize=.235\hsize\linewidth=\hsize\centering\arraybackslash}X | >{\hsize=.145\hsize\linewidth=\hsize\centering\arraybackslash}X | >{\hsize=.145\hsize\linewidth=\hsize\centering\arraybackslash}X |>{\hsize=.145\hsize\linewidth=\hsize\centering\arraybackslash}X | >{\hsize=.145\hsize\linewidth=\hsize\centering\arraybackslash}X | >{\hsize=.15\hsize\linewidth=\hsize\centering\arraybackslash}X |>{\hsize=.15\hsize\linewidth=\hsize\centering\arraybackslash}X | >{\hsize=.15\hsize\linewidth=\hsize\centering\arraybackslash}X | >{\hsize=.15\hsize\linewidth=\hsize\centering\arraybackslash}X |>{\hsize=.15\hsize\linewidth=\hsize\centering\arraybackslash}X | >{\hsize=.145\hsize\linewidth=\hsize\centering\arraybackslash}X | >{\hsize=.145\hsize\linewidth=\hsize\centering\arraybackslash}X |>{\hsize=.145\hsize\linewidth=\hsize\centering\arraybackslash}X }
    \hline
    & \multicolumn{2}{c|}{Gender} & \multicolumn{2}{|c|}{Education}   & \multicolumn{3}{|c|}{Rent [€]}    & \multicolumn{3}{|c|}{City Center Distance}& \multicolumn{2}{|c}{Car Available}    \\\hline
    Groups & F   & M & Non-uni. & Uni.    &   <500   & <700  &   <900 &   <1km   &   <5km  &   <10km   &   Yes &   No    \\\hline
    Eff. [s]   &    147.3   &   145.1   &    116.8  &   147.3   &   141.0   &   145.7   &   144.2   &  164.2  &  112.1  &  182.1 & 163.5  & 127.8 \\\hline
    Effect.  &  0.5    &   0.2   &   0.6    &   -0.4   &  -0.2  & 0.5 & 0.3 & -0.3 & 0.5  &  0.1 &  0.3 & 0.2  \\\hline
    Sat.    & 3.8 &   3.6 &   3.8 &   3.5   &  4  &  3.5  &  3.6 & 3.9 & 3.7  & 3.5  &  3.6 &  3.8  \\\hline
    Transp.    & 3.8   &   3.9  &  4.1  &  3.5 &  4.2  & 3.9  &  3.5  & 4.3 &  4.0 &  3.3  &  4.0 &  3.7 \\\hline    
    \end{tabularx}
    \caption{\textbf{High-involvement} domain (apartments), grouped by demographics and preferences (excluding \textit{children} due to even distribution).}
    \label{tab:apartment_demographics}
\end{subtable}
\raggedbottom
\end{table}

\raggedbottom

\subsubsection{Effectiveness and Persuasiveness}
To measure effectiveness, the ratings for the explanation-based and content-description-based recommendations of the same user are compared \cite{bilgic2005explaining,tintarev2012evaluating}. Similar ratings indicate effectiveness, as both ways of presenting the suggestion led to the same result. If the full content description of a recommendation receives lower ratings than the explanation, users tend to overestimate the explanation, indicating persuasiveness. On the contrary case, users tend to underestimate the explanation, indicating weak persuasiveness. 

The descriptive statistics (see Table \ref{tab:rating_statistics}) show that all observed explanation types, including the baseline, were effective, with minimal mean rating differences. This is expected as the necessary information to decide is already included in the explanation. However, motivating consequences showed a trend of slight persuasiveness, while avoiding consequences and the baseline tended to be underestimated. Consequence-based explanations were effective for both low- and high-involvement item domains. For apartments, users found motivating consequences slightly more persuasive, resembling advertising tactics that highlight positive aspects. For recipes, users tended to underestimate the item, particularly when avoiding consequences were mentioned, likely because negative consequences are generally weak in this domain.

The analysis suggests that consequence-based explanations provide an effective way to explain recommendations in both low- and high-involvement item domains, where the motivating formulation leads to slight persuasiveness.

\subsubsection{Satisfaction} 
To assess user satisfaction with the explanations, participants were asked to rate their satisfaction with the recommended item \cite{tintarev2012evaluating,gedikli2014should}. Consequence-based explanations resulted in higher user satisfaction compared to the baseline in both low- and high-involvement domains. Users appreciated the clear communication of the suggested item's impact, with avoidance of consequences leading to the highest user satisfaction. Aligned with the \textit{prospect theory} \cite{kahneman1979prospect}, it indicates that users tend to prioritize avoiding negative impacts when making choices based on the explanation.

Based on the descriptive statistics (see Table \ref{tab:rating_statistics}), we found that user satisfaction was higher in the high-involvement domain than in the low-involvement domain, which supports our assumption and may be explained due to the higher criticality of the decision's impact. While no difference was observed between motivating and avoiding consequences in the apartment domain, avoiding consequences received higher ratings in the recipe domain. Overall, the results indicate that consequence-based explanations are well-received by users and positively affect user satisfaction with the RS.

\subsubsection{Transparency}
To evaluate the \textit{user-perceived transparency} of the explanations \cite{tintarev2012evaluating,gedikli2014should}, participants rated how well the explanation helped them understand the reasoning behind the recommendation. The descriptive statistics indicate (see Table \ref{tab:rating_statistics}) that consequence-based explanations, in both formulations and domains, did not excel in terms of transparency as they lack insights into how the system generated the suggestion.

\subsubsection{Summary}
In response to RQ1, RQ2, and RQ3, the study found that consequence-based explanations are efficient for explaining recommendations in both domains. Motivating consequences were more efficient in the low-involvement domain, while mentioning avoided negative consequences was more efficient in the high-involvement domain. Both formulations were effective, with motivating consequences being slightly more persuasive than avoiding consequences. Users expressed higher satisfaction compared to content-based explanations, with the avoiding formulation leading to the highest satisfaction. In terms of transparency, consequence-based explanations did not outperform the baseline.

\subsection{Influence of Demographics and User Preferences}
RQ4 explores the connection between users' demographics, preferences, and their impact on the explanation aims. Descriptive statistics for both domains are presented in Tables \ref{tab:recipe_demographics} and \ref{tab:apartment_demographics}. The analysis excluded age due to participant homogeneity. Persons with higher education levels spent more time analyzing explanations in both domains, likely due to their tendency for thorough decision-making. In the recipe domain, participants with specific requirements (e.g., preferred cooking time, weight aim) evaluated explanations faster and found them more transparent, showing a better understanding of the recommendation process.
Female participants, those without a preference for cooking time, and very active individuals tended to underestimate the recommended recipe. This suggests that more detailed and persuasive explanations could be beneficial for those. For apartments, participants without a university degree found the explanations more persuasive, while graduates tended to underestimate the recommended item. This indicates that graduates may prefer more detailed and comprehensible explanations. Conclusions for user satisfaction are not drawn due to homogeneous results.

\subsection{Importance of Features}
To determine which features in consequences influence users' decisions to consume a recommended item, participants rated their importance. Analyzing the data using the \textit{Kruskal-Wallis} test showed significant differences across the features in both domains, which have been identified with pairwise \textit{Mann-Whitney-U} tests. 
The results indicate that consequences with a strong personal impact were valued higher than those describing the item more generally. In the apartment domain, \textit{financial liability} is an important factor to be included, with the \textit{monthly rate} consequence being more important than the \textit{distance to the city center}, a \textit{private garden}, a \textit{private parking spot}, \textit{number of bedrooms}, and \textit{distance to preferred leisure activities}. Furthermore, the \textit{number of bedrooms} and the \textit{distance to the city center} are more important than having a \textit{garden}. In the recipe domain, \textit{preparation time} is more important than the \textit{cuisine type}, \textit{difficulty level} of a recipe, and individual \textit{activity level}. The \textit{weight aim}, \textit{preferred diet}, and \textit{avoidance of ingredients} are more important than the \textit{activity level} consequence. 

\section{Conclusions, Limitations, and Future Work} \label{sec:conclusion_and_future_work}
This paper introduced and evaluated consequence-based explanations as a new explanation type for recommendations, emphasizing the individual impact of a recommended item. Our user study confirmed user appreciation, showing a trend toward increased satisfaction. These explanations support users in making effective and efficient decisions, particularly in high-involvement domains where the impact is crucial, and users may lack expertise. By highlighting personal impact, valuable insights for understanding the reasons behind recommendations and making informed decisions are provided.

The initial study analyzes this novel explanation type, while it acknowledges certain limitations that will be addressed in future work. One limitation is the lack of statistically significant results for some measured dimensions. A reason might be the small number of participants in the study. To overcome this, we plan to conduct a larger study with a more diverse group of participants, in terms of age, gender, and educational background, to gain a deeper understanding of the findings. Another limitation is the use of synthetic datasets, which may impact the evaluation of recommended items. To address this concern, we intend to validate the consequence-based explanations in a real-world study, which will provide practical insights. Finally, the current methods for generating our explanations are limited to the recommended item content. In the future, we aim to enhance this method by incorporating collaborative data and refining personalization techniques. Additionally, we expect the integration of generative AI to offer diverse and natural explanations, ensuring their authenticity and eliminating any false consequences, as a beneficial possibility for improving the explanations.

\begin{acknowledgments}
The presented work has been developed within the research project \textsc{Streamdiver} which is funded by the Austrian Research Promotion Agency (FFG) under the project number \textsc{886205}.
\end{acknowledgments}

\bibliography{bibliography}

\begin{thebibliography}{24}
\expandafter\ifx\csname natexlab\endcsname\relax\def\natexlab#1{#1}\fi
\providecommand{\url}[1]{\texttt{#1}}
\providecommand{\href}[2]{#2}
\providecommand{\path}[1]{#1}
\providecommand{\DOIprefix}{doi:}
\providecommand{\ArXivprefix}{arXiv:}
\providecommand{\URLprefix}{URL: }
\providecommand{\Pubmedprefix}{pmid:}
\providecommand{\doi}[1]{\href{http://dx.doi.org/#1}{\path{#1}}}
\providecommand{\Pubmed}[1]{\href{pmid:#1}{\path{#1}}}
\providecommand{\bibinfo}[2]{#2}
\ifx\xfnm\relax \def\xfnm[#1]{\unskip,\space#1}\fi
\bibitem[{Ricci et~al.(2022)Ricci, Rokach, and Shapira}]{ricci2022recommender}
\bibinfo{author}{F.~Ricci}, \bibinfo{author}{L.~Rokach},
  \bibinfo{author}{B.~Shapira}, \bibinfo{title}{Recommender Systems:
  Techniques, Applications, and Challenges}, \bibinfo{publisher}{Springer US},
  \bibinfo{address}{New York, NY}, \bibinfo{year}{2022}, pp.
  \bibinfo{pages}{1--35}. \DOIprefix\doi{10.1007/978-1-0716-2197-4_1}.
\bibitem[{Gomez-Uribe and Hunt(2016)}]{gomez-uribe2015netflix}
\bibinfo{author}{C.~A. Gomez-Uribe}, \bibinfo{author}{N.~Hunt},
\newblock \bibinfo{title}{The netflix recommender system: Algorithms, business
  value, and innovation},
\newblock \bibinfo{journal}{ACM Trans. Manage. Inf. Syst.} \bibinfo{volume}{6}
  (\bibinfo{year}{2016}). \DOIprefix\doi{10.1145/2843948}.
\bibitem[{Sharma and Dutta(2020)}]{sharma2020movie}
\bibinfo{author}{N.~Sharma}, \bibinfo{author}{M.~Dutta},
\newblock \bibinfo{title}{Movie recommendation systems: A brief overview},
\newblock in: \bibinfo{booktitle}{Proceedings of the 8th International
  Conference on Computer and Communications Management}, ICCCM '20,
  \bibinfo{publisher}{Association for Computing Machinery},
  \bibinfo{address}{New York, NY, USA}, \bibinfo{year}{2020}, p.
  \bibinfo{pages}{59–62}. \DOIprefix\doi{10.1145/3411174.3411194}.
\bibitem[{Ge et~al.(2015)Ge, Ricci, and Massimo}]{ge2015health}
\bibinfo{author}{M.~Ge}, \bibinfo{author}{F.~Ricci},
  \bibinfo{author}{D.~Massimo},
\newblock \bibinfo{title}{Health-aware food recommender system},
\newblock in: \bibinfo{booktitle}{Proceedings of the 9th ACM Conference on
  Recommender Systems}, RecSys '15, \bibinfo{publisher}{Association for
  Computing Machinery}, \bibinfo{address}{New York, NY, USA},
  \bibinfo{year}{2015}, p. \bibinfo{pages}{333–334}.
  \DOIprefix\doi{10.1145/2792838.2796554}.
\bibitem[{Pecune et~al.(2020)Pecune, Callebert, and
  Marsella}]{pecune2020recommender}
\bibinfo{author}{F.~Pecune}, \bibinfo{author}{L.~Callebert},
  \bibinfo{author}{S.~Marsella},
\newblock \bibinfo{title}{A recommender system for healthy and personalized
  recipe recommendations},
\newblock \bibinfo{year}{2020}.
\bibitem[{Burke(2000)}]{burke2000knowledge}
\bibinfo{author}{R.~Burke},
\newblock \bibinfo{title}{Knowledge-based recommender systems},
\newblock \bibinfo{journal}{Encyclopedia of library and information systems}
  \bibinfo{volume}{69} (\bibinfo{year}{2000}) \bibinfo{pages}{175--186}.
\bibitem[{Gharahighehi et~al.(2021)Gharahighehi, Pliakos, and
  Vens}]{gharahighehi2021recommender}
\bibinfo{author}{A.~Gharahighehi}, \bibinfo{author}{K.~Pliakos},
  \bibinfo{author}{C.~Vens},
\newblock \bibinfo{title}{Recommender systems in the real estate market—a
  survey},
\newblock \bibinfo{journal}{Applied Sciences} \bibinfo{volume}{11}
  (\bibinfo{year}{2021}). \DOIprefix\doi{10.3390/app11167502}.
\bibitem[{Bilgic and Mooney(2005)}]{bilgic2005explaining}
\bibinfo{author}{M.~Bilgic}, \bibinfo{author}{R.~J. Mooney},
\newblock \bibinfo{title}{Explaining recommendations: Satisfaction vs.
  promotion},
\newblock in: \bibinfo{booktitle}{Beyond personalization workshop, IUI},
  volume~\bibinfo{volume}{5}, \bibinfo{year}{2005}, p. \bibinfo{pages}{153}.
\bibitem[{Zhang and Chen(2020)}]{zhang2020explainable}
\bibinfo{author}{Y.~Zhang}, \bibinfo{author}{X.~Chen},
\newblock \bibinfo{title}{Explainable recommendation: A survey and new
  perspectives},
\newblock \bibinfo{journal}{Found. Trends Inf. Retr.} \bibinfo{volume}{14}
  (\bibinfo{year}{2020}) \bibinfo{pages}{1–101}.
  \DOIprefix\doi{10.1561/1500000066}.
\bibitem[{Felfernig et~al.(2017)Felfernig, Atas, Tran, Stettinger, Erdeniz, and
  Leitner}]{felfernig2017analysis}
\bibinfo{author}{A.~Felfernig}, \bibinfo{author}{M.~Atas},
  \bibinfo{author}{T.~N.~T. Tran}, \bibinfo{author}{M.~Stettinger},
  \bibinfo{author}{S.~P. Erdeniz}, \bibinfo{author}{G.~Leitner},
\newblock \bibinfo{title}{An analysis of group recommendation heuristics for
  high-and low-involvement items},
\newblock in: \bibinfo{booktitle}{Advances in Artificial Intelligence: From
  Theory to Practice: 30th International Conference on Industrial Engineering
  and Other Applications of Applied Intelligent Systems, IEA/AIE 2017, Arras,
  France, June 27-30, 2017, Proceedings, Part I 30},
  \bibinfo{organization}{Springer}, \bibinfo{year}{2017}, pp.
  \bibinfo{pages}{335--344}.
\bibitem[{Tintarev and Masthoff(2007)}]{tintarev2007survey}
\bibinfo{author}{N.~Tintarev}, \bibinfo{author}{J.~Masthoff},
\newblock \bibinfo{title}{A survey of explanations in recommender systems},
\newblock in: \bibinfo{booktitle}{2007 IEEE 23rd international conference on
  data engineering workshop}, \bibinfo{organization}{IEEE},
  \bibinfo{year}{2007}, pp. \bibinfo{pages}{801--810}.
\bibitem[{Gedikli et~al.(2014)Gedikli, Jannach, and Ge}]{gedikli2014should}
\bibinfo{author}{F.~Gedikli}, \bibinfo{author}{D.~Jannach},
  \bibinfo{author}{M.~Ge},
\newblock \bibinfo{title}{How should i explain? a comparison of different
  explanation types for recommender systems},
\newblock \bibinfo{journal}{International Journal of Human-Computer Studies}
  \bibinfo{volume}{72} (\bibinfo{year}{2014}) \bibinfo{pages}{367--382}.
\bibitem[{Gkika and Lekakos(2014)}]{gkika2014persuasive}
\bibinfo{author}{S.~Gkika}, \bibinfo{author}{G.~Lekakos},
\newblock \bibinfo{title}{The persuasive role of explanations in recommender
  systems.},
\newblock in: \bibinfo{booktitle}{BCSS@ PERSUASIVE}, \bibinfo{year}{2014}, pp.
  \bibinfo{pages}{59--68}.
\bibitem[{Zanker and Ninaus(2010)}]{zanker2010knowledgable}
\bibinfo{author}{M.~Zanker}, \bibinfo{author}{D.~Ninaus},
\newblock \bibinfo{title}{Knowledgeable explanations for recommender systems},
\newblock in: \bibinfo{booktitle}{2010 IEEE/WIC/ACM International Conference on
  Web Intelligence and Intelligent Agent Technology},
  volume~\bibinfo{volume}{1}, \bibinfo{year}{2010}, pp.
  \bibinfo{pages}{657--660}. \DOIprefix\doi{10.1109/WI-IAT.2010.131}.
\bibitem[{Tran et~al.(2021)Tran, Le, Atas, Felfernig, Stettinger, and
  Popescu}]{tran2021users}
\bibinfo{author}{T.~N.~T. Tran}, \bibinfo{author}{V.~M. Le},
  \bibinfo{author}{M.~Atas}, \bibinfo{author}{A.~Felfernig},
  \bibinfo{author}{M.~Stettinger}, \bibinfo{author}{A.~Popescu},
\newblock \bibinfo{title}{Do users appreciate explanations of recommendations?
  an analysis in the movie domain},
\newblock in: \bibinfo{booktitle}{Fifteenth ACM Conference on Recommender
  Systems}, \bibinfo{year}{2021}, pp. \bibinfo{pages}{645--650}.
\bibitem[{Tintarev and Masthoff(2022)}]{tintarev2022beyond}
\bibinfo{author}{N.~Tintarev}, \bibinfo{author}{J.~Masthoff},
  \bibinfo{title}{Beyond Explaining Single Item Recommendations},
  \bibinfo{publisher}{Springer US}, \bibinfo{address}{New York, NY},
  \bibinfo{year}{2022}, pp. \bibinfo{pages}{711--756}.
  \DOIprefix\doi{10.1007/978-1-0716-2197-4_19}.
\bibitem[{{van der Waa} et~al.(2018){van der Waa}, {van Diggelen}, {van den
  Bosch}, and {Neerincx}}]{vanderwaa2018contrastive}
\bibinfo{author}{J.~{van der Waa}}, \bibinfo{author}{J.~{van Diggelen}},
  \bibinfo{author}{K.~{van den Bosch}}, \bibinfo{author}{M.~{Neerincx}},
\newblock \bibinfo{title}{{Contrastive Explanations for Reinforcement Learning
  in terms of Expected Consequences}},
\newblock \bibinfo{journal}{arXiv e-prints}  (\bibinfo{year}{2018}).
  \DOIprefix\doi{10.48550/arXiv.1807.08706}.
\bibitem[{Ripberger et~al.(2015)Ripberger, Silva, Jenkins-Smith, and
  James}]{ripberger_2015_influence}
\bibinfo{author}{J.~T. Ripberger}, \bibinfo{author}{C.~L. Silva},
  \bibinfo{author}{H.~C. Jenkins-Smith}, \bibinfo{author}{M.~James},
\newblock \bibinfo{title}{The influence of consequence-based messages on public
  responses to tornado warnings},
\newblock \bibinfo{journal}{Bulletin of the American Meteorological Society}
  \bibinfo{volume}{96} (\bibinfo{year}{2015}) \bibinfo{pages}{577--590}.
\bibitem[{Fano and Kurth(2003)}]{fano2003personal}
\bibinfo{author}{A.~E. Fano}, \bibinfo{author}{S.~W. Kurth},
\newblock \bibinfo{title}{Personal choice point: helping users visualize what
  it means to buy a bmw},
\newblock in: \bibinfo{booktitle}{International Conference on Intelligent User
  Interfaces}, \bibinfo{year}{2003}.
\bibitem[{Miller(2019)}]{miller2019explanation}
\bibinfo{author}{T.~Miller},
\newblock \bibinfo{title}{Explanation in artificial intelligence: Insights from
  the social sciences},
\newblock \bibinfo{journal}{Artificial Intelligence} \bibinfo{volume}{267}
  (\bibinfo{year}{2019}) \bibinfo{pages}{1--38}.
  \DOIprefix\doi{10.1016/j.artint.2018.07.007}.
\bibitem[{Davidson et~al.(2010)Davidson, Livingston, Sampath, Liebald, Liu,
  Nandy, Van~Vleet, Gargi, Gupta, He, and Lambert}]{davidson2010youtube}
\bibinfo{author}{J.~Davidson}, \bibinfo{author}{B.~Livingston},
  \bibinfo{author}{D.~Sampath}, \bibinfo{author}{B.~Liebald},
  \bibinfo{author}{J.~Liu}, \bibinfo{author}{P.~Nandy},
  \bibinfo{author}{T.~Van~Vleet}, \bibinfo{author}{U.~Gargi},
  \bibinfo{author}{S.~Gupta}, \bibinfo{author}{Y.~He},
  \bibinfo{author}{M.~Lambert},
\newblock \bibinfo{title}{The {YouTube} video recommendation system},
\newblock in: \bibinfo{booktitle}{Proceedings of the fourth {ACM} conference on
  {Recommender} systems - {RecSys} '10}, \bibinfo{publisher}{ACM Press},
  \bibinfo{address}{Barcelona, Spain}, \bibinfo{year}{2010}, p.
  \bibinfo{pages}{293}. \DOIprefix\doi{10.1145/1864708.1864770}.
\bibitem[{Jansen(2011)}]{jansen2011maut}
\bibinfo{author}{S.~Jansen}, \bibinfo{title}{The Multi-attribute Utility
  Method}, \bibinfo{year}{2011}, pp. \bibinfo{pages}{101--125}.
  \DOIprefix\doi{10.1007/978-90-481-8894-9_5}.
\bibitem[{Kahneman and Tversky(1979)}]{kahneman1979prospect}
\bibinfo{author}{D.~Kahneman}, \bibinfo{author}{A.~Tversky},
\newblock \bibinfo{title}{Prospect theory: An analysis of decision under risk},
\newblock \bibinfo{journal}{Econometrica} \bibinfo{volume}{47}
  (\bibinfo{year}{1979}) \bibinfo{pages}{263--291}.
\bibitem[{Tintarev and Masthoff(2012)}]{tintarev2012evaluating}
\bibinfo{author}{N.~Tintarev}, \bibinfo{author}{J.~Masthoff},
\newblock \bibinfo{title}{Evaluating the effectiveness of explanations for
  recommender systems: Methodological issues and empirical studies on the
  impact of personalization},
\newblock \bibinfo{journal}{User Modeling and User-Adapted Interaction}
  \bibinfo{volume}{22} (\bibinfo{year}{2012}) \bibinfo{pages}{399--439}.

\end{thebibliography}


\end{document}